\documentclass[11pt,a4paper]{article}
\usepackage{epsfig,wrapfig}
\newcommand{\be}{\begin{equation}}
\newcommand{\ee}{\end{equation}}
\newcommand{\bea}{\begin{eqnarray}}
\newcommand{\eea}{\end{eqnarray}}

\begin{document}
\begin{titlepage}

\begin{flushright}
WUE-ITP-2003-038\\
LAPTH-1021/03\\
December 2003
\end{flushright}
\vspace{1.cm}

\begin{center}
\large\bf
{\LARGE\bf A compact representation of the $\gamma\gamma ggg\rightarrow 0$ amplitude}\\[1cm]
\rm
{ T.~Binoth$^{a}$, J.~Ph.~Guillet$^{b}$, F.~Mahmoudi$^{b}$}\\[1cm]

{\em $^{a}$Institut f\"ur Theoretische Physik und Astrophysik\\
           Universit\"at W\"urzburg\\
	   Am Hubland\\
	   D-97074 W\"urzburg, Germany} \\[.5cm]

{\em $^{b}$LAPTH\\
           Chemin de Bellevue B.P. 110\\ 
	   F-74941 Annecy-le-Vieux, France} \\[.5cm]

\end{center}
\normalsize

\vspace{1cm}

\begin{abstract}
A compact representation of the loop amplitude $\gamma\gamma ggg\rightarrow 0$
is presented. The result has been obtained by using helicity methods and 
sorting with respect to an irreducible function basis. We show how to convert 
spinor representations into a field strength representation of the amplitude. 
The amplitude defines a background contribution for Higgs
boson searches at the LHC in the channel $H \to \gamma\gamma + jet$ which was earlier
extracted indirectly from the one-loop representation of the 5-gluon amplitude.

\end{abstract}



\end{titlepage}

\section{Introduction}

Collider experiments at the TeV scale and especially the forthcoming LHC 
experiment at CERN produce a large amount of data
which has to be cross checked with theoretical predictions. To understand 
signatures from the
Standard Model and beyond on a quantitative level, a good understanding
of signal {\em and} background reactions is mandatory. 
This fact motivates a considerable effort from the theory side
to describe prominent signal and background reactions beyond tree level.
This task is highly challenging due to the combinatorial complexity
of the Feynman diagrammatic approach if the number of external particles increases.
Even the number of known 5-point 1-loop amplitudes is very restricted.
Notable exceptions relevant for multi-jet \cite{Bern:1993mq,Kunszt:1994tq,Bern:1997sc,Yasui:2002bn} 
and Higgs physics exist 
\cite{Reina:2001bc,Beenakker:2002nc,Beenakker:2001rj,Belanger:2003ya,Belanger:2002ik,Belanger:2003nm}. 
Up to now not a single Standard Model process which has generic  $2\to 4$ 
kinematics is computed at the one-loop level although this is relevant
for many search channels at the LHC.
Examples for such amplitudes are provided only for the Yukawa model 
\cite{Binoth:2002qh,Binoth:2001vm} or for special helicity configurations 
\cite{Mahlon:1993si,Mahlon:1993fe,Bern:1993qk,Bern:1994cg,Bern:zx,Bern:1998xc}.

To attack such a highly involved problem new methods have to be investigated
and efficient, constructive algorithms have to be developed which allow
an automated computation of the corresponding amplitudes. 
To test and investigate new ideas and concepts we calculate the 
$gg \to \gamma\gamma g$  amplitude in the present paper.

In hadronic collisions this amplitude is relevant for the production of photon pairs
in association with a jet and as such a contribution of the background
to the Higgs boson search channel $H\to\gamma\gamma + \mbox{jet}$.
A phenomenological analysis has already been provided in \cite{deFlorian:1999tp,DelDuca:2003uz}.
Further this amplitude contributes to the 2-loop corrections for $gg \to\gamma\gamma$ \cite{Bern:2002jx}.
In both articles the  $gg \to \gamma\gamma g$ amplitude has been extracted from the well known 1-loop  5-gluon 
amplitude \cite{Bern:1993mq} by replacing two gluons by photons and modifying the colour algebra
of the original result.  
No direct computation has been presented yet.
We will see that a compact representation for this amplitude can be obtained
by decomposing the amplitude in helicity components and  
by using an irreducible function basis.  

In the next section we will explain the organization of our calculation 
and the used methods. Especially we will show how
spinorial representations can be retranslated into 
field strength representations of the result and
give some rules to switch between equivalent representations. 
In section 3 our result will be presented in terms
of helicity amplitudes which are expressed by field strength tensors.   
Section 4 will close the paper with a summary and an outlook 
for future applications of our methods.

\section{Algebraic methods}

We will outline here briefly the basic technology we have used to perform the
calculation. We will focus here predominantly on features which we think
are not standard and will not provide all details of the given calculation
in this article.

\subsection{Preliminaries}

We consider here the one-loop amplitude for two photon three gluon scattering, $\gamma\gamma ggg \to 0$.
It is defined by QED and QCD Feynman rules.
The amplitude with all momenta in-going has the following
kinematics:
\begin{equation}
\gamma(p_1,\lambda_1) + \gamma(p_2,\lambda_2) 
+ g( p_3,\lambda_3,c_3 ) + g( p_4,\lambda_4,c_4 ) + g( p_5,\lambda_5,c_5 ) \to 0
\end{equation}
$\lambda_j$ and $c_l$ are helicity and  the colour indices.
The conversion to physical kinematics is done by crossing rules. 

\begin{figure}[ht]
\begin{picture}(150,40)
\put(30,0){\epsfig{file=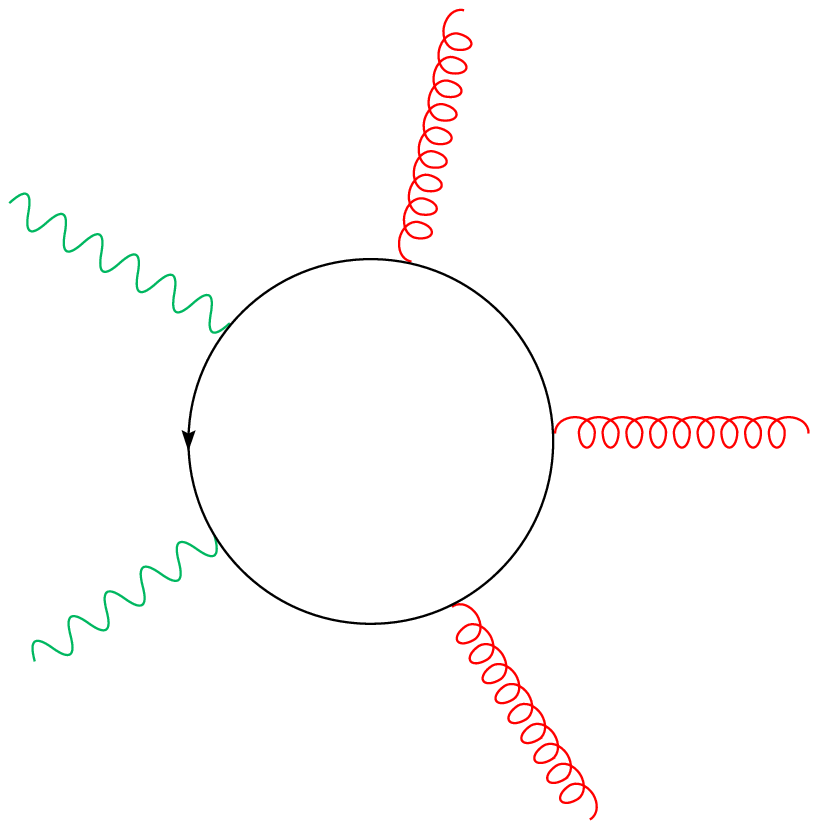,height=4.cm}}
\put(80,5){\epsfig{file=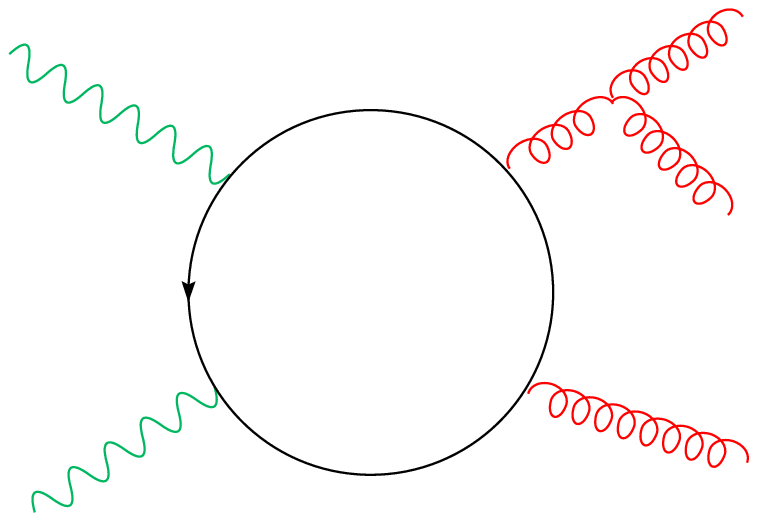,height=2.5cm}}
\end{picture}
\caption{\em The two basic topologies of the $\gamma\gamma ggg$ amplitude. All the other
graphs are obtained by reordering of photons and gluons.}
\label{topologies}
\end{figure}

The amplitude consists of 24 pentagon Feynman graphs
and 18 box diagrams\footnote{The triangle diagrams with a 4-gluon vertex
vanish due to colour conservation.}. In  Fig.\ref{topologies} the two basic 
topologies are shown. 

The colour structure of the pentagon topology is proportional to
a trace of three colour matrices  $T^{c}$ in the fundamental representation.
With the normalization ${\rm tr}(T^aT^b)=\delta_{ab}/2$ one has:
${\rm tr}(T^{c_3}T^{c_4}T^{c_5}) = ( d^{c_3c_4c_5}+if^{c_3c_4c_5} )/4$. 
Because of Furry's theorem only
the totally antisymmetric contribution in colour space
remains which is also present in the 4-point topology, such that
there is only one colour structure.   
Subsequently the amplitude will be written as
\begin{equation}
\Gamma^{\{\lambda_j\},\{c_j\}}[\gamma\gamma g g g \to 0] 
= \frac{Q_q^2 g_s^3}{i \pi^2} f^{c_3c_4c_5} {\cal A}^{\lambda_1\lambda_2\lambda_3\lambda_4\lambda_5}
\end{equation}
All numeric factors are shifted to ${\cal A}^{\lambda_1\lambda_2\lambda_3\lambda_4\lambda_5}$.
In principle there are 32 helicity amplitudes but some of them are related by
parity conservation. One has
\begin{eqnarray}\label{hel_amps}
&& {\cal A}^{+++++}(1,2,3,4,5) = {\cal A}^{-----}(1,2,3,4,5) \nonumber\\
&& {\cal A}^{++++-}(1,2,3,4,5) = {\cal A}^{----+}(1,2,3,4,5) \nonumber\\
&& {\cal A}^{-++++}(1,2,3,4,5) = {\cal A}^{+----}(1,2,3,4,5) \nonumber\\
&& {\cal A}^{--+++}(1,2,3,4,5) = {\cal A}^{++---}(1,2,3,4,5) \nonumber\\
&& {\cal A}^{+++--}(1,2,3,4,5) = {\cal A}^{---++}(1,2,3,4,5) \nonumber\\
&& {\cal A}^{-+++-}(1,2,3,4,5) = {\cal A}^{+---+}(1,2,3,4,5) 
\end{eqnarray}
All the others are obtained by interchange of the photons and/or 
cyclic relabeling of the gluons. Thus, it is sufficient to calculate 
the 6 helicity amplitudes in (\ref{hel_amps}).
Note that because of Bose symmetry and the  colour structure of the amplitude 
anti-symmetry relations for the auxiliary objects ${\cal A}$ exist.  
For example:
\begin{equation}
{\cal A}^{+++--}(1,2,3,4,5)= -{\cal A}^{+++--}(1,2,3,5,4) = -{\cal A}^{++-+-}(1,2,4,3,5)
\end{equation}
The Bose symmetry allows for stringent checks of the calculation.


\subsection{Spinor helicity methods}
The use of helicity methods for loop calculations is well established 
meanwhile \cite{Xu:1986xb,Bern:1996je}.
The polarization vectors
are represented by spinor products
\bea
\epsilon^{+}_{\mu}(p,k) &=& \frac{<k^-|\mu|p^->}{\sqrt{2}<k^-|p^+>} \nonumber \\
\epsilon^{-}_{\mu}(p,k) &=& \frac{<k^+|\mu|p^+>}{\sqrt{2}[pk]}\quad,\quad [pk]= <k^-|p^+>^*
\eea
Here $k$ is an arbitrary light-like vector defining the transverse direction for the
respective photon/gluon polarization vector. 

To perform the calculation we followed two strategies. In a first approach we 
constructed an algebraic program where the reference vectors can be chosen arbitrarily.  
In  a second approach we have chosen the reference momenta for the 
polarization vectors  such that
there is a global spinorial factor for all Feynman diagrams. 
In this way a spinorial factor can be extracted from the amplitude
and all graphs can be treated uniformly for a given helicity amplitude.

To give an example for the second approach
let us consider the amplitude $\Gamma^{-+++-}$.
After computing the traces of the one-loop integrals
the numerator of the integrals is a polynomial
of scalar products of polarization vectors and four momenta, $P(\epsilon_i\cdot \epsilon_j,
\epsilon_i\cdot p_j,p_i\cdot p_j)$.    
Products of
polarization vectors are expressible in terms of 
scalar products between polarization vectors and momenta by using the following formulas
\begin{eqnarray}\label{convEEtoEP}
\epsilon^+(i,j)\cdot \epsilon^+(k,l) &=& \left\{
\begin{array}{ll}
0 & \mbox{if} \;\; j=l \\
\frac{1}{p_i\cdot p_l} \;\; \epsilon^+(i,j)\cdot p_l \;\; \epsilon^+(k,l)\cdot p_i 
& \mbox{if} \; i\neq l, j\neq l \\
\frac{1}{p_j\cdot p_k} \;\; \epsilon^+(i,j)\cdot p_k\;\; \epsilon^+(k,l)\cdot p_j 
& \mbox{if} \;j\neq k,i = l, j\neq l \\
-\frac{p_i\cdot p_j}{p_i\cdot p_m \; p_j\cdot p_m} \epsilon^+(i,j)\cdot p_k \;\; \epsilon^+(k,l)\cdot p_j &\mbox{if} \;\; j\neq k,i = l, j\neq l \\
\end{array}\right. \nonumber \\ && \nonumber \\
\epsilon^+(i,j)\cdot \epsilon^-(k,l) &=& \left\{
\begin{array}{ll}
0 & \qquad\quad \mbox{if} \;\;i=l\; \mbox{or}\; j=k \\
\frac{1}{p_i\cdot p_k} \; \epsilon^+(i,j)\cdot p_k \;\;\epsilon^-(k,l)\cdot p_i &\qquad\quad\mbox{if} \;\; i\neq l, j \neq k
\end{array} \right.
\end{eqnarray}
The arguments stand for the respective index of the external momentum.
By applying these formulas to the numerator polynomials $P$, each numerator is now 
a linear combination of terms $\sim \prod\limits_{j=1}^{5} \epsilon_j\cdot q_j$.
Using for example the reference momenta $(p_5,p_5,p_2,p_3,p_1)$ for the polarization
vectors $\epsilon_1^-,\epsilon_2^+,\epsilon_3^+,\epsilon_4^+,\epsilon_5^-$ and formulas (\ref{convEEtoEP})
one can express these products by  spinor traces through
\begin{eqnarray}
\prod \limits_{j=1}^{5} \epsilon_j\cdot q_j &=& 
   \frac{<5^-| q_5 1 q_1 5 q_2 2 q_3 3 q_4 |4^-> }{4\sqrt{2}[15]<52><23><34>[51]}\nonumber\\
&=&\frac{\mbox{tr}^-(5q_51q_1 5 q_2 2 q_3 3 q_4 4 2)}{4\sqrt{2}[15]<52><23><34>[51]<42>[25]}
\end{eqnarray}
Here we use the notations ${\rm tr}^\pm(j,\dots)={\rm tr}^\pm(p_j,\dots)= {\rm tr}([1\pm\gamma_5]p_j,\dots)/2$
and $<j^-|l^+>=<jl>$.
To achieve the trace representation a factor one, $1=<4^-|2|5^->/<4^-|2|5^->$, was included. 
Here the vector $p_2$ may be replaced by any other light-like vector.  
The spinor product is now a global factor multiplying the given helicity amplitude and does not enter
the subsequent reduction algebra of scalar and tensor integrals.
 
After tensor and scalar integral reduction one finds a result of the form
\begin{eqnarray}
{\cal A}^{-+++-} = \frac{A + \varepsilon(1,2,3,4)\, B}{4\sqrt{2}[15]<52><23><34>[51]<42>[25]}
\end{eqnarray} 
$A$ and $B$ stand for combinations of Mandelstam variables and scalar integrals.
The epsilon tensor here is defined by $\varepsilon(i,j,k,l)={\rm tr}(\gamma_5 \,i\, j \,k\, l)$.
For the given reference momenta one has: 
$\epsilon_1\cdot p_5=\epsilon_2\cdot p_5=\epsilon_3\cdot p_2=\epsilon_4\cdot p_3=\epsilon_5\cdot p_1=0$.
Using this fact the spinor representation can be rewritten in terms of field strength tensors, 
${\cal F}^{\mu\nu}_j=p^\mu_j\epsilon^\nu_j-p^\nu_j\epsilon^\mu_j$.  The non-Abelian part of the gluon
field strength tensor
does not enter in the given order in $\alpha_s$.
By rewriting the epsilon tensor in terms of spinors and identifying the spinor products with 
polarization vectors and external vectors one arrives at
the field strength representation of the given helicity amplitude. With the abbreviations
$C_a=[ A+B\,\mbox{tr}(2354) ]$, $C_b=-2 s_{23}s_{24} B$ one gets:
\begin{eqnarray}
{\cal A}^{-+++-}_F &=& {\rm Tr}({\cal F}_1^-{\cal F}_5^-){\rm Tr}({\cal F}_2^+{\cal F}_3^+)
\Bigl[ C_a\;  p_3\cdot {\cal F}_4^+ \cdot p_2  + C_b\;  p_3\cdot {\cal F}_4^+ \cdot p_5 \Bigr] 
\end{eqnarray}
The field strength objects are more explicitly
\begin{eqnarray}
{\rm Tr}({\cal F}_i^\pm{\cal F}_j^\pm) &=&  
2 \, p_i\cdot \epsilon_j^\pm p_j\cdot \epsilon_i^\pm - s_{ij}\, \epsilon_i^\pm\cdot \epsilon_j^\pm \nonumber\\
p_i\cdot {\cal F}_j^\pm \cdot p_k &=& ( s_{ij}\, p_k\cdot \epsilon_j^\pm - s_{jk}\, p_i\cdot \epsilon_j^\pm )/2
\end{eqnarray}
The Mandelstam variables are defined by $s_{ij}=2 p_i\cdot p_j$.
The obtained representations are not unique and we will discuss below how to
transform the result into equivalent representations.
An analogous approach is possible  for all the other helicity amplitudes, too.
We will see below that all amplitudes are constructed from traces and
so-called tails \cite{Schubert:2001he}, which are Lorentz invariants 
built out of field strength
tensors and external momenta. Here only the most simple type for a tail term appears, 
$p_j\cdot {\cal F}_k^\pm \cdot p_l$. 

Using helicity methods in amplitude calculations typically leads 
to expressions where spinor products are the basic 
building blocks. Within the  method presented here complex phases do not appear
in the final results, which is of advantage concerning numerical evaluation. 

\subsection{Basic analytical structures}
As explained in the last subsection each Feynman diagram is a linear combination
of spinor terms $<i^-|\dots k\dots p_l\dots|j^->$. It was already observed elsewhere
\cite{Pittau:1996ez,Pittau:1997mv,Weinzierl:1998we} that it is always possible to convert such an expression to
a form with inverse propagators and a spinor trace which contains
at most one power of the loop momentum. This means that only rank 1 5-point tensor integrals
have to be reduced which simplifies the calculation considerably. 
It is easy to see  that this statement holds true for a general massive 
5-point amplitude too, by using the helicity formalism for massive particles 
and applying the same reasoning as outlined above. A similar observation was made
recently in \cite{Denner:2002ii}. We want to remark that an equivalent statement is true for general 
$N$-point amplitudes.

Note that by power counting all 5-point graphs are IR and UV finite.
The tensor reduction introduces spurious UV and IR divergences.  The box topologies
are UV divergent but the divergence cancels in the sum over all box graphs. 
For the reduction of the rank 1 5-point function and the 4-point tensor integrals
we applied the formalism outlined in \cite{Binoth:1999sp,Heinrich:2000kj}.
After tensor reduction one is left with scalar integrals which are reduced
to a basic set of integrals using well known formulas 
\cite{Binoth:1999sp,Bern:1992em}. We have used FORM \cite{Vermaseren:2000nd} 
to perform the tensor and scalar integral reductions.

\begin{figure}[ht]
\begin{picture}(150,15)
\put(10,0){\epsfig{file=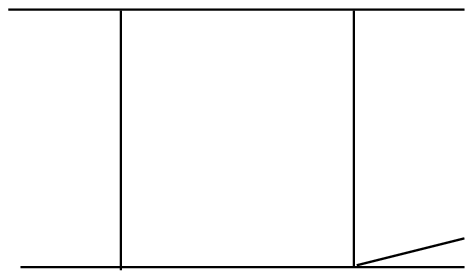,height=1.5cm}}
\put(45,0){\epsfig{file=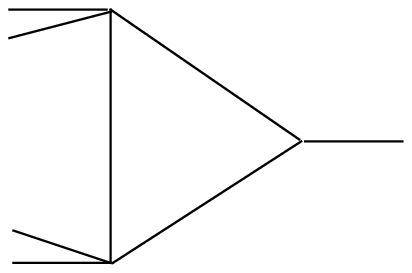,height=1.5cm}}
\put(80,0){\epsfig{file=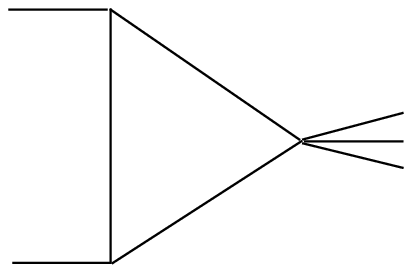,height=1.5cm}}
\put(115,0){\epsfig{file=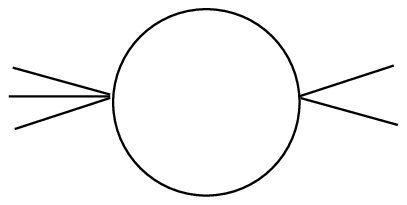,height=1.5cm}}
\end{picture}
\caption{\em The scalar graph topologies of the $\gamma\gamma ggg$ amplitude.}
\label{scalar_basis}
\end{figure}

The 5-point functions are reduced to box integrals with one external massive line. 
The latter are reduced to triangle functions and box functions in $(n+2)$ dimensions,
where $n=4-2\epsilon$. 
In this way the spurious infrared divergences   are isolated in a transparent way.
For a given ordering of the external legs, say (1,2,3,4,5), one 
finds the following basis of scalar functions, see also Fig.\ref{scalar_basis}:
\begin{eqnarray}
&& I_4^{n+2}(p_1,p_2,p_3,p_4+p_5) + \mbox{4 cyclic permutations}\nonumber\\
&& I_3^n(p_1,p_2+p_3,p_4+p_5) + \mbox{4 cyclic permutations} \nonumber\\
&& I_3^n(p_1,p_2,p_3+p_4+p_5) + \mbox{4 cyclic permutations} \nonumber\\
&& I_2^n(p_1+p_2,p_3+p_4+p_5) + \mbox{4 cyclic permutations} \nonumber
\end{eqnarray}
We are following the notation in \cite{Binoth:1999sp}.
In addition the amplitude contains a finite part from the ultra-violet region of
the one loop integrals. Terms of order ${\cal O}(n-4)$ combine with UV divergent
2-point integrals $I_2^n\sim 1/\epsilon$ to constant terms.
The coefficients of these
functions are first simplified on a graph by graph basis.
As each graph is infrared finite the triangle  graphs with one
and two external legs off-shell cancel as a whole. No expansion
in $\epsilon$ is performed for the scalar integrals. 
The different analyticity properties, or in other words, the different cut 
structure  of each basis function does not allow any cross talking between
the coefficients. It follows that each graph is expressible
by 6-dimensional box integrals, 2-point functions and a constant term.
For later use we define the following dimensionless function:
\be
F_1(s_{j_1j_2},s_{j_2j_3},s_{j_4j_5}) = 
\frac{1}{s_{j_4j_5}-s_{j_1j_2}-s_{j_2j_3}} \; I_4^6(p_{j_1},p_{j_2},p_{j_3},p_{j_4}+p_{j_5})
\ee
Its analytic representation is well known and can be found in \cite{Binoth:2001vm}.
Our result will be expressed entirely in terms of this function, the scalar 2-point integral in $n=4-2\epsilon$
dimensions,
\be
I_2^n(s_{ij}) = \frac{\Gamma(1+\epsilon)\Gamma(1-\epsilon)^2}{\Gamma(2-2\epsilon)} \frac{(-s_{ij})^{-\epsilon}}{\epsilon}
\ee
and a constant term. 
As the amplitude is UV finite also the sum of coefficients
of the 2-point functions have to cancel.
For each helicity amplitude the coefficient of a given
function was first simplified for each graph. Then the sum over all graphs
was performed and the result again simplified using MAPLE \cite{maple}. 
The application of the helicity method together 
with the sorting of a given diagram into a function bases proved to be an efficient 
method for the calculation of the given amplitude.
The size of the expressions was never a problem in our approach as
we have split up the calculation as much as possible into
irreducible building blocks.  

\subsection{Rules for field strength objects}
We have shown above that one can convert spinor representations into field strength 
representations of the amplitudes. The latter are not unique 
and we quote two useful formulas to relate equivalent representations
to each other. The first is a permutation rule:
\bea\label{perm_rule}
{\rm Tr}({\cal F}_1^+{\cal F}_2^+) \; p_i\cdot {\cal F}_3^+\cdot p_j &=& \frac{{\rm Tr}({\cal F}_1^+{\cal F}_3^+)}{s_{13}^2s_{23}}
\Bigl\{
\Bigl[ s_{23}( s_{1j}s_{3i} - s_{1i} s_{3j} ) + s_{13} (s_{2i}s_{3j} - s_{2j}s_{3i})  \Bigr] p_3\cdot {\cal F}_2^+\cdot p_1\nonumber\\
&&+ s_{12}s_{13} ( s_{3i} \; p_3\cdot {\cal F}_2^+\cdot p_j - s_{3j} \; p_3\cdot {\cal F}_2^+\cdot p_i )
\Bigr\}
\eea
Of course the formula is true for arbitrary indices. The indices 1,2,3 were taken for convenience.
In a given helicity basis it is easy to see that for a given field strength tensor  only two 
tails, $p_i\cdot {\cal F} \cdot p_j$, are independent. This is directly related to the fact that 
there are only two independent scalar products between a polarization vector and external vectors. 
   
The second useful identity is a flipping rule:
\bea\label{flip_rule}
{\rm Tr}({\cal F}_1^+{\cal F}_2^+) {\rm Tr}({\cal F}_3^+{\cal F}_4^+) &=& 
{\rm Tr}({\cal F}_1^+{\cal F}_3^+){\rm Tr}({\cal F}_2^+{\cal F}_4^+)\left( \frac{{\rm tr}^+(1243)}{s_{13}s_{24}} \right)^2\nonumber\\
                              &=& \frac{{\rm Tr}({\cal F}_1^+{\cal F}_3^+)}{s_{13}^2}\frac{{\rm Tr}({\cal F}_2^+{\cal F}_4^+)}{s_{24}^2}
			      [ {\rm tr}(1243){\rm tr}^+(1243) -s_{12}s_{13}s_{34}s_{24} ]
\eea
Both identities can be verified by choosing an arbitrary set of reference momenta
and the identities in Eq.~(\ref{convEEtoEP}). By complex conjugation, $(\epsilon_j^+)^*=\epsilon_j^-$,
the formulas hold also for negative helicities. As ${\rm Tr}({\cal F}_i^+{\cal F}_j^-)=0$, no flipping or permutation
rules exist between different helicities.

\section{Result}

We present now our result by expressing each helicity amplitude through field strength
tensors and coefficients for the functions $F_1,I_2^n,1$. The Ward identities
are manifest in this representation.

\subsection{${\cal A}^{+++++}$}

For the helicity amplitude with all helicities plus, all 6-dimensional box and 2-point functions cancel.
This can be understood by the fact that every amplitude is defined, up to a polynomial
remainder, by the sum over all cuts \cite{Bern:1994cg}. Each cut  corresponds to the gluing
of two tree level amplitudes, in our case 2 or 3 gauge bosons attached to a fermion line. 
The tree level amplitude of a massless fermion
and two/three gauge bosons with equal helicities vanishes. It follows that 
no non-polynomial function, i.e. $F_1$ or $I_2^n$, can be present  
in the amplitude. 
By choosing reference momenta in a cyclic way, $p_j$ for $\epsilon_{j+1}$, we find a term of the form
\begin{eqnarray}
{\cal A}^{+++++} = \frac{1}{2\sqrt{2}<51><12><23><34><45>}\mbox{tr}^+(1,2,3,5)\frac{s_{12}}{s_{35}}
\end{eqnarray}
which can be written as 
\begin{eqnarray}
{\cal A}^{+++++} = \frac{ \mbox{Tr}({\cal F}_1^+{\cal F}_2^+)\mbox{Tr}({\cal F}_3^+{\cal F}_4^+{\cal F}_5^+) }{2\,s_{34}s_{45}s_{35} }
\end{eqnarray}
where $F_j^{\mu\nu} = p_j^\mu \epsilon_j^\nu - p_j^\nu\epsilon_j^\mu$ is the Abelian part of the field strength tensor
of the photons and gluons. In this form Ward identities are manifest. The expression is symmetric under exchange of
the two photons and antisymmetric under gluon exchange. Including the antisymmetric 
colour factor leads to the full $S_2 \otimes S_3$ Bose symmetry of this amplitude.
\subsection{${\cal A}^{-++++}$}
The only allowed structure is a polynomial. We find
\begin{eqnarray}
{\cal A}^{-++++} = \frac{\mbox{Tr}({\cal F}_2^+{\cal F}_3^+)\mbox{Tr}({\cal F}_4^+{\cal F}_5^+) }{s_{23}^2 s_{45}^2}
\Bigl[ C^{-++++}\;  p_2\cdot {\cal F}_1^- \cdot p_4 - ( 4 \leftrightarrow 5 ) \Bigr]
\end{eqnarray} with the coefficient 
\begin{equation}
 C^{-++++} = -\frac{s_{15}s_{12}}{s_{24}s_{35}} 
                        -\frac{s_{15}}{s_{35}} + \frac{s_{23}}{s_{24}} - \frac{s_{15}}{s_{34}}
\end{equation}
Using the shift rules one finds that ${\cal A}^{-++++}$ is totally anti-symmetric in the gluon indices 3,4,5.
Taking into account the colour factor this leads to the full $S_3$ Bose symmetry of the
corresponding helicity amplitude.

\subsection{${\cal A}^{++++-}$}

Again no nontrivial function can be present due to cutting rules.
For the polynomial term we find the following result
\begin{eqnarray}
{\cal A}^{++++-} = \frac{\mbox{Tr}({\cal F}_1^+{\cal F}_2^+)\mbox{Tr}({\cal F}_3^+{\cal F}_4^+)}{s_{12}^2 s_{34}^2} 
\Bigl[ C^{++++-}  \; p_1\cdot {\cal F}_5^- \cdot p_3 - ( 3 \leftrightarrow 4 ) \Bigr]
\end{eqnarray} 

The coefficient is given by
\begin{eqnarray}
C^{++++-} &=& -\frac{s_{45}s_{13}s_{14}}{s_{35}s_{15}s_{24}} 
-\frac{s_{13}s_{45}}{s_{15}s_{35}}
+\frac{s_{45}^{2}}{s_{15}s_{24}}
-\frac{s_{12}^{2}+s_{45}^{2}-s_{12}s_{45}}{s_{35}s_{15}}
+\frac{s_{13}s_{15}}{s_{23}s_{45}}
+\frac{s_{13}-s_{34}}{s_{23}} \nonumber\\ &&
-\frac{s_{34}s_{45}}{s_{23}s_{15}}
+\frac{s_{15}-s_{25}}{s_{45}}
-\frac{s_{23}+s_{35}}{s_{13}}
-\frac{s_{23}s_{25}}{s_{13}s_{45}}
+\frac{s_{34}+s_{12}}{s_{15}}
\end{eqnarray}
We have checked that the corresponding amplitude has a $S_2\otimes S_2$ Bose symmetry when the photons
and the gluons with equal helicities are interchanged.

\subsection{${\cal A}^{--+++}$}
Now a nontrivial cut structure is possible. The cuts which can be associated with
the Mandelstam variables $s_{13},s_{14},s_{15},s_{23},s_{24},s_{25}$ are allowed by the
helicity structure. No other Mandelstam variable is to be expected as an argument
of the functions $F_1,I_2^n$ and indeed is not observed.  
We split the result into three pieces with indices $F,B,1$, which belong to the  
 part proportional to 6-dimensional 
boxes $F_1$,  a part containing bubble graphs $I_2^n$,  and a constant term, respectively.   
\begin{eqnarray}
{\cal A}^{--+++} = {\cal A}^{--+++}_F + {\cal A}^{--+++}_B + {\cal A}^{--+++}_1 
\end{eqnarray} 
We find

\begin{eqnarray}
{\cal A}^{--+++}_F &=& \frac{\mbox{Tr}({\cal F}_1^-{\cal F}_2^-)\mbox{Tr}({\cal F}_3^+{\cal F}_4^+)}{s_{12}^2 s_{34}^2} 
\Bigl[ C^{--+++}_F\; p_1\cdot {\cal F}_5^+\cdot p_3 - (  3 \leftrightarrow 4 ) \Bigr] F_1(s_{13},s_{14},s_{25})\nonumber\\ &&
- (  4 \leftrightarrow 5 ) - (  5 \leftrightarrow 3 ) + (  1 \leftrightarrow 2 )  
       - (  1 \leftrightarrow 2, 4 \leftrightarrow 5 ) - (  1 \leftrightarrow 2, 5 \leftrightarrow 3 )\nonumber\\
{\cal A}^{--+++}_B &=& \frac{\mbox{Tr}({\cal F}_1^-{\cal F}_2^-)\mbox{Tr}({\cal F}_3^+{\cal F}_4^+)}{s_{12}^2 s_{34}^2} 
\Bigl[ C^{--+++}_B\; p_1\cdot {\cal F}_5^+\cdot p_3 - (  3 \leftrightarrow 4 ) \Bigr] I_2^n(s_{15})\nonumber\\ &&
- (  4 \leftrightarrow 5 ) - (  5 \leftrightarrow 3 ) + (  1 \leftrightarrow 2 )  
       - (  1 \leftrightarrow 2, 4 \leftrightarrow 5 ) - (  1 \leftrightarrow 2, 5 \leftrightarrow 3 )\nonumber\\       
{\cal A}^{--+++}_1 &=& \frac{\mbox{Tr}({\cal F}_1^-{\cal F}_2^-)\mbox{Tr}({\cal F}_3^+{\cal F}_4^+{\cal F}_5^+)}{2\,s_{34}s_{45}s_{s35}}
\end{eqnarray} 
The indicated permutations have to be applied to the respective initial term
for the indices of the field strength tensors, momenta and  Mandelstam variables. 
The coefficients are
\begin{eqnarray}
C^{--+++}_F &=& \frac{1}{2}\frac{s_{12}^2-2s_{13}s_{14}}{s_{35}s_{15}} 
                         - \frac{s_{14}}{s_{34}}- \frac{s_{14}}{s_{35}}\nonumber\\
C^{--+++}_B &=& \frac{s_{45}}{s_{15}}\left[\frac{s_{13}+s_{35}}{s_{14}+s_{45}}+\frac{s_{14}+s_{45}}{s_{13}+s_{35}}\right]
                         +\frac{s_{45}^2s_{13}}{s_{15} s_{35}^2}
			 +\frac{s_{14}s_{35}}{s_{15} s_{45}}+2\frac{(s_{15}+s_{45})^2}{s_{35}^2}
			 \nonumber\\ &&
                        -\frac{s_{13}+s_{35}}{s_{15}}
			-\frac{s_{14}s_{45}}{s_{15} s_{35}}
			-\frac{s_{45}^2}{s_{35} s_{15}}
			+\frac{s_{14}+s_{24}}{s_{45}}
			+\frac{s_{12}-s_{14}-s_{35}}{s_{14}+s_{45}}
			+2\frac{s_{14}(s_{15}+s_{45})}{s_{35}^2}
			\nonumber\\ &&
			+\frac{s_{23}^2s_{15}}{s_{35}^2(s_{13}+s_{35})}
			+\frac{2s_{45}+s_{15}}{s_{13}+s_{35}}
			-2\frac{(s_{15}+s_{45})s_{23}}{s_{35}(s_{13}+s_{35})}
			-\frac{(2s_{45}+s_{15})}{s_{35}}
			+\frac{s_{13}(2s_{45}+s_{15})}{s_{35}^2}
			\nonumber
\end{eqnarray} 
In the given expressions the  $S_2 \otimes S_3$ symmetry 
under exchange of the two photons and the three gluons is manifest 
after taking into account the omitted colour factor.
We note that we have actually calculated all coefficients of the $F_1,I_2^n$ functions and checked the symmetry
explicitly. A further nontrivial check is provided by the fact that the sum of all coefficients of the bubble
integrals add up to zero. Of course the amplitude is finite and the $1/\epsilon$ poles cancel. 
To show this the flipping and shifting rules (\ref{perm_rule},\ref{flip_rule}) for the field strength
expressions have to be applied.
The logarithmic terms can be organized as logarithms of ratios of Mandelstam variables. 
The presence of the constant term can be inferred from the $+++++$ amplitude. Gauge invariance predicts
this term as a part of the result. The fact that no other constant term is present seems to be an 
effect of the high symmetry of this amplitude. 

\subsection{${\cal A}^{+++--}$}
The amplitude has a $S_2\otimes S_2$ symmetry under exchange of the gluons with equal helicities
and the photons. Although we use this symmetry in the representation we have calculated all 
the coefficients of the functions independently and checked the symmetries explicitly.
Again we write
\begin{eqnarray}
{\cal A}^{+++--} = {\cal A}^{+++--}_F + {\cal A}^{+++--}_B + {\cal A}^{+++--}_1 
\end{eqnarray}  
The amplitude has  less symmetries than ${\cal A}^{--+++}$ above:  
\begin{eqnarray}
{\cal A}^{+++--}_F &=&\frac{\mbox{Tr}({\cal F}_1^+{\cal F}_2^+)\mbox{Tr}({\cal F}_4^-{\cal F}_5^-)}{s_{12}^2 s_{45}^2}
\Bigl[ [ C^{+++--}_{F\,1}\;  p_1\cdot {\cal F}_3^+ \cdot p_4 + ( 1 \leftrightarrow 2 )] F_1(s_{14},s_{24},s_{35}) 
  - ( 4 \leftrightarrow 5 ) \Bigr] \nonumber \\
&& 
+\frac{\mbox{Tr}({\cal F}_4^-{\cal F}_5^-)}{s_{45}^2}\Bigl[ \frac{\mbox{Tr}({\cal F}_1^+{\cal F}_3^+)}{s_{13}^2} [C^{+++--}_{F\,2}\;  p_1\cdot {\cal F}_2^+ \cdot p_4  - ( 1 \leftrightarrow 3 )] 
   \; F_1(s_{14},s_{34},s_{25}) \nonumber\\&& \hspace{4cm}+ ( 1 \leftrightarrow 2 ) - ( 4 \leftrightarrow 5 ) 
   - ( 1 \leftrightarrow 2,4 \leftrightarrow 5 )\Bigr] \nonumber    
\end{eqnarray} 
The coefficients are

\begin{eqnarray}
 C^{+++--}_{F\,1} &=&
\Bigl[ 
 -\frac{2s_{24}^{2}s_{34}}{s_{12}^{2}s_{35}}-\frac{2s_{24}(s_{23}+s_{24})}{s_{12}^{2}}-\frac{3s_{24}}{s_{12}}+\frac{2s_{24}(-s_{34}+2s_{24})}{s_{35}s_{12}}
\nonumber \\ &&    
 -\frac{(s_{14}^{2}+s_{24}^{2})}{2s_{13}s_{34}}-\frac{(s_{14}^{2}+s_{24}^{2})}{s_{34}s_{35}}-\frac{(2s_{14}-2s_{24})}{s_{35}}-\frac{s_{34}}{s_{35}}-\frac{(s_{34}+2s_{14})}{2s_{13}}\Bigr]
\nonumber\\ 
 C^{+++--}_{F\,2} &=&
\Bigl[\frac{s_{34}^{2}}{2s_{12}s_{24}}+\frac{(s_{34}+s_{45})^{2}}{2s_{12}s_{24}}-\frac{s_{34}}{s_{13}}\Bigr]
\nonumber\end{eqnarray}

The nonvanishing two-point functions are
\begin{eqnarray}
{\cal A}^{+++--}_B &=&\frac{\mbox{Tr}({\cal F}_1^+{\cal F}_2^+)\mbox{Tr}({\cal F}_4^-{\cal F}_5^-)}{s_{12}^2 s_{45}^2}
\Bigl[ [ C^{+++--}_{B\,1}\;  p_1\cdot {\cal F}_3^+ \cdot p_4 + ( 1 \leftrightarrow 2 )] I_2^n(s_{34}) 
  - ( 4 \leftrightarrow 5 ) \Bigr] \nonumber \\
&& 
+\frac{\mbox{Tr}({\cal F}_4^-{\cal F}_5^-)}{s_{45}^2}
\Bigl[ 
   \frac{\mbox{Tr}({\cal F}_1^+{\cal F}_3^+)}{s_{13}^2} [C^{+++--}_{B\,2a}\;  p_1\cdot {\cal F}_2^+ \cdot p_4  
                                                           + C^{+++--}_{B\,2b}\;  p_3\cdot {\cal F}_2^+ \cdot p_4 ] 
   \; I_2^n(s_{15}) \nonumber\\&& \hspace{4cm}+ ( 1 \leftrightarrow 2 )  - ( 4 \leftrightarrow 5 ) 
   - ( 1 \leftrightarrow 2,4 \leftrightarrow 5 )\Bigr] \nonumber    
\end{eqnarray} 

with the coefficients
\begin{eqnarray}
&& C^{+++--}_{B\,1} =
\Bigl[
-\frac{s_{12}s_{24}^{2}}{s_{34}(s_{23}+s_{24})^{2}}
-\frac{s_{24}(s_{12}-s_{24})}{(s_{23}+s_{24})^{2}}
+\frac{s_{24}(s_{45}+2s_{24}+s_{12})}{s_{34}(s_{23}+s_{24})}
-\frac{2s_{24}^{2}}{s_{12}(s_{23}+s_{24})}
\nonumber\\&&
+\frac{s_{12}}{(s_{23}+s_{24})}
-\frac{s_{23}(3s_{14}+s_{12}+s_{23})}{s_{34}s_{13}}
-\frac{s_{23}^{2}s_{14}}{s_{34}s_{13}^{2}}
+\frac{s_{23}(s_{23}+s_{12}+s_{14})}{s_{34}(s_{13}+s_{14})}
-\frac{s_{23}s_{14}s_{12}}{s_{34}(s_{13}+s_{14})^{2}}
\nonumber\\&&
+\frac{(s_{13}+s_{23})(-s_{23}+s_{15})}{s_{34}s_{23}}
+\frac{s_{14}s_{34}(s_{12}-s_{14})^{2}}{s_{13}^{2}(s_{13}+s_{14})^{2}}
-\frac{s_{23}s_{14}^{2}(-s_{14}+4s_{12})}{s_{12}s_{13}^{2}(s_{13}+s_{14})}
-\frac{s_{23}s_{14}^{2}(2s_{12}-s_{14})}{s_{13}^{2}(s_{13}+s_{14})^{2}}
\nonumber\\&&
-\frac{2s_{14}s_{25}}{s_{12}(s_{13}+s_{14})}
-\frac{2s_{12}s_{34}(-2s_{14}+s_{12})}{s_{13}^{2}(s_{13}+s_{14})}
-\frac{3s_{14}^{3}s_{23}}{s_{13}^{2}s_{12}(s_{13}+s_{14})}
+\frac{s_{23}s_{14}(4s_{12}+5s_{14})}{s_{13}^{2}(s_{13}+s_{14})}
-\frac{2s_{14}^{2}s_{23}}{s_{13}^{2}s_{12}}
\nonumber\\&&
-\frac{2s_{23}s_{14}(s_{13}-2s_{14})}{s_{13}^{2}s_{12}}
+\frac{(2s_{23}+2s_{12}+2s_{34})}{s_{13}}
-\frac{(s_{34}s_{14}+2s_{12}s_{34}+2s_{23}s_{14})}{s_{13}^{2}}
-\frac{(s_{12}+s_{13}-s_{23})}{s_{23}}\Bigr]
\nonumber\end{eqnarray}

\begin{eqnarray}
&& C^{+++--}_{B\,2a} =\Bigl[
\frac{s_{34}^{2}s_{24}^{3}}{(s_{23}+s_{24})^{2}s_{23}s_{12}^{2}}
-\frac{s_{34}s_{24}^{2}(-s_{24}+3s_{34}+2s_{13})}{s_{12}(s_{23}+s_{24})^{2}s_{23}}
+\frac{s_{24}(-2s_{24}+3s_{34}+2s_{13})s_{34}}{(s_{23}+s_{24})^{2}s_{23}}
\nonumber\\&&
-\frac{s_{34}(6s_{23}-3s_{34}+s_{24}-2s_{45})}{s_{12}s_{23}}
+\frac{2s_{34}^{2}s_{12}}{s_{23}s_{24}(s_{23}+s_{24})}
-\frac{7s_{34}}{(s_{23}+s_{24})}
-\frac{s_{34}s_{24}(s_{34}+3s_{13}-s_{24})}{(s_{23}+s_{24})s_{12}^{2}}
\nonumber\\&&
+\frac{s_{12}s_{34}(s_{24}-s_{34})}{(s_{23}+s_{24})^{2}s_{23}}
-\frac{s_{13}s_{34}(s_{34}+s_{24})}{s_{23}s_{24}(s_{23}+s_{34})}
-\frac{s_{45}^{2}}{s_{12}^{2}}
-\frac{2s_{23}^{2}(3s_{34}s_{45}+s_{34}^{2}+3s_{45}^{2})}{s_{24}s_{12}^{3}}
-\frac{s_{34}}{(s_{23}+s_{34})}
\nonumber\\&&
-\frac{2s_{24}s_{34}^{2}}{s_{12}^{3}}
-\frac{s_{24}s_{34}^{2}}{s_{12}^{2}s_{23}}
-\frac{s_{34}(2s_{23}-3s_{34}-3s_{45}+s_{24})}{s_{12}^{2}}
+\frac{2s_{23}(-3s_{45}^{2}+s_{34}^{2})}{s_{12}^{3}}
-\frac{2s_{45}^{3}(2s_{23}+s_{24})}{s_{12}^{3}s_{34}}
\nonumber\\&&
+\frac{s_{34}(-2s_{45}+2s_{12}-3s_{34})}{s_{24}s_{23}}
+\frac{2s_{34}^{2}s_{24}^{2}}{(s_{23}+s_{24})s_{12}^{3}}
-\frac{s_{45}^{3}s_{23}}{s_{12}^{2}s_{34}s_{24}}
+\frac{s_{34}(7s_{34}+5s_{13}+3s_{12})}{(s_{23}+s_{24})s_{24}}
+\frac{5s_{34}}{s_{24}}
\nonumber\\&&
-\frac{2s_{23}^{2}s_{45}^{3}}{s_{12}^{3}s_{34}s_{24}}
-\frac{s_{23}(4s_{34}^{2}+6s_{45}^{2}+9s_{34}s_{45})}{s_{12}^{2}s_{24}}
-\frac{s_{34}(6s_{34}+4s_{13}-3s_{24})}{(s_{23}+s_{24})s_{12}}
+\frac{s_{45}^{2}(2s_{23}+2s_{24}-s_{45}+s_{12})}{s_{12}^{2}s_{34}}
\Bigr]
\nonumber\end{eqnarray}

\begin{eqnarray}
&& C^{+++--}_{B\,2b} =\Bigl[
\frac{s_{35}^{2}}{(s_{23}+s_{24})^{2}}
-\frac{s_{34}(2s_{35}+s_{12})}{(s_{23}+s_{24})^{2}}
+\frac{s_{24}(s_{24}+2s_{34}-2s_{35})}{(s_{23}+s_{24})^{2}}
-\frac{(2s_{12}s_{34}+s_{13}^{2})}{s_{23}^{2}}
\nonumber\\&&
+\frac{2s_{14}s_{34}s_{23}}{(s_{23}+s_{24})s_{12}^{2}}
-\frac{2s_{14}s_{34}s_{23}}{s_{24}s_{12}^{2}}
-\frac{2s_{12}s_{34}(s_{34}+s_{12})}{s_{24}s_{23}^{2}}
-\frac{2s_{35}s_{34}(s_{24}+s_{12})}{(s_{23}+s_{34})s_{23}^{2}}
-\frac{4s_{14}s_{23}s_{45}}{s_{24}s_{12}^{2}}
\nonumber\\&&
-\frac{2s_{14}}{(s_{23}+s_{24})}
-\frac{s_{12}s_{34}(-3s_{35}+s_{12})}{s_{24}(s_{23}+s_{24})^{2}}
-\frac{s_{12}s_{34}^{2}s_{35}}{(s_{23}+s_{34})s_{23}^{2}s_{24}}
-\frac{s_{14}(s_{14}+s_{24})^{2}}{s_{12}s_{34}s_{24}}
-\frac{(2s_{13}+s_{23})}{s_{23}}
\nonumber\\&&
+\frac{s_{34}s_{24}s_{14}}{s_{12}(s_{23}+s_{24})^{2}}
-\frac{s_{14}s_{12}s_{34}}{s_{24}^{2}(s_{23}+s_{24})}
-\frac{s_{14}s_{12}s_{34}}{s_{24}s_{23}^{2}}
+\frac{s_{14}s_{12}s_{34}}{s_{24}^{2}s_{23}}
-\frac{2s_{14}s_{45}^{2}s_{23}}{s_{12}^{2}s_{34}s_{24}}
-\frac{(2s_{14}-s_{34})s_{14}}{(s_{23}+s_{24})s_{12}}
\nonumber\\&&
+\frac{s_{12}^{2}s_{34}^{2}}{s_{24}^{2}(s_{23}+s_{24})^{2}}
-\frac{s_{12}^{2}s_{34}^{2}}{s_{23}^{2}s_{24}^{2}}
-\frac{s_{34}^{2}(s_{24}+s_{12})}{(s_{23}+s_{34})s_{23}^{2}}
+\frac{s_{14}(3s_{14}+s_{24})}{s_{12}s_{24}}
-\frac{s_{24}s_{35}^{2}}{(s_{23}+s_{34})s_{23}^{2}}
-\frac{2s_{14}s_{45}^{2}}{s_{34}s_{12}^{2}}
\Bigr]
\nonumber\end{eqnarray}

The constant term is
\begin{eqnarray}
{\cal A}^{+++--}_1 &=&\frac{\mbox{Tr}({\cal F}_1^+{\cal F}_2^+)\mbox{Tr}({\cal F}_4^-{\cal F}_5^-)}{s_{12}^2 s_{45}^2} 
[ C^{+++--}_{1}\;  p_2\cdot {\cal F}_3^+ \cdot p_4  - ( 4 \leftrightarrow 5 )]    
\end{eqnarray} 

with the coefficient

\begin{eqnarray}
&& C^{+++--}_1 =\nonumber \\ && \Bigl[
-\frac{s_{12}(s_{15}+2s_{13}+s_{14})}{s_{23}(-s_{34}+s_{15})}
+\frac{(s_{15}+s_{35})(s_{23}+s_{15})s_{25}}{(-s_{34}+s_{15})s_{23}s_{34}}
+\frac{s_{14}(2s_{15}+s_{14})}{s_{13}(s_{13}+s_{14})}
-\frac{(s_{13}-s_{15})^{2}(s_{34}+s_{13})}{(s_{13}+s_{14})s_{23}s_{14}}\nonumber \\ &&
+\frac{(-s_{14}+s_{13})^{2}(s_{34}+s_{13})}{s_{13}s_{23}(s_{15}+s_{13})}
+\frac{s_{14}(s_{35}s_{14}-s_{15}^{2})}{s_{13}s_{35}(s_{15}+s_{13})}
-\frac{s_{15}^{2}s_{14}}{s_{13}s_{23}s_{35}}
+\frac{s_{14}^{2}}{s_{23}s_{13}}
+\frac{s_{23}s_{13}}{s_{14}(s_{14}-s_{35})}\nonumber \\ &&
+\frac{s_{23}s_{15}}{s_{35}(s_{14}-s_{35})}
+\frac{s_{35}(-s_{24}+2s_{23})}{s_{14}(s_{14}-s_{35})}
-\frac{(s_{13}+s_{14})s_{25}}{s_{35}s_{14}}
+\frac{(s_{13}+2s_{12})}{s_{34}}
-\frac{s_{14}s_{15}(s_{13}+s_{34}-s_{15})}{s_{13}s_{35}s_{23}}\nonumber \\ &&
+\frac{(s_{12}^{2}+2s_{12}s_{13}+s_{13}^{2}+s_{12}s_{25}+s_{25}s_{13}+s_{25}^{2})}{s_{23}s_{34}}
 +\frac{(s_{14}-s_{15})}{s_{13}}
 -\frac{(s_{13}-s_{34}+s_{15}-3s_{23})}{s_{14}}\nonumber \\ &&
 +\frac{(s_{13}-s_{15})^{2}(s_{34}+s_{13})}{s_{23}s_{13}s_{14}}
 -\frac{(s_{15}+s_{14})(-3s_{34}+s_{14})}{s_{23}s_{13}}
 -\frac{(-3s_{35}+s_{13}-3s_{15}-2s_{14})}{s_{23}}
\Bigr]\nonumber\end{eqnarray}

\subsection{${\cal A}^{-+++-}$}

The amplitude has only a $S_2$ symmetry under exchange of the gluons with equal helicities.
Again we split the amplitude into an $F_1$, $I_2^n$ and constant part. 
\begin{eqnarray}
{\cal A}^{-+++-} = {\cal A}^{-+++-}_F + {\cal A}^{-+++-}_B + {\cal A}^{-+++-}_1 
\end{eqnarray} 
with
\begin{eqnarray}
{\cal A}^{-+++-}_F &=&
\Bigl\{ \Bigl[ [ C^{-+++-}_{F\,1}\;  p_1\cdot {\cal F}_2^+ \cdot p_3 \;F_1(s_{13},s_{14},s_{25}) ]\nonumber\\ 
&&  \;\;  + [ C^{-+++-}_{F\,2}\;  p_1\cdot {\cal F}_2^+ \cdot p_3 \;F_1(s_{35},s_{45},s_{12}) ]\Bigr]
\frac{\mbox{Tr}({\cal F}_1^-{\cal F}_5^-)\mbox{Tr}({\cal F}_3^+{\cal F}_4^+)}{s_{15}^2 s_{34}^2}\nonumber\\   
&& + \Bigl[[ C^{-+++-}_{F\,3}\;  p_1\cdot {\cal F}_4^+ \cdot p_2  + ( 2 \leftrightarrow 3 ) ]\;F_1(s_{12},s_{13},s_{45})\nonumber\\  
&& \;\; + [ C^{-+++-}_{F\,4a}\;  p_1\cdot {\cal F}_4^+ \cdot p_2 + C^{-+++-}_{F\,4b}\;  p_1\cdot {\cal F}_4^+ \cdot p_3  ]\;F_1(s_{25},s_{35},s_{14})\Bigr]
\nonumber\\ &&\qquad \frac{\mbox{Tr}({\cal F}_1^-{\cal F}_5^-)\mbox{Tr}({\cal F}_2^+{\cal F}_3^+)}{s_{15}^2 s_{23}^2}\Bigl\} \;\;- \;\;( 3 \leftrightarrow 4 )\nonumber
\end{eqnarray} 
and
\begin{eqnarray}
C^{-+++-}_{F\,1} &=& \frac{1}{2}\frac{s_{15}^2}{s_{12}s_{23}} - \frac{s_{13}s_{14}}{s_{12}s_{23}} 
                                                  - \frac{s_{14}}{s_{34}}         - \frac{s_{14}}{s_{23}}\nonumber\\ 
C^{-+++-}_{F\,2} &=& -C^{-+++-}_{F\,1} \nonumber\\
C^{-+++-}_{F\,3} &=&-\frac{1}{2}\frac{s_{15}^2}{s_{14}s_{24}} + \frac{s_{12}s_{13}}{s_{14}s_{24}} 
                + \frac{s_{13}}{s_{24}}  + \frac{s_{13}}{s_{23}} -\frac{s_{15}^2-2s_{12}s_{13}}{s_{14}s_{45}}	\nonumber\\&&
				 + 2\frac{s_{13}}{s_{23}}\Bigl[  1+\frac{s_{24}}{s_{23}}-\frac{s_{12}}{s_{45}}
				    -\frac{s_{12}s_{15}}{s_{23}s_{45}} \Bigr] \nonumber\\	   
C^{-+++-}_{F\,4a} &=& \frac{1}{2}\frac{s_{15}^2-2s_{12}s_{13} }{s_{14}s_{24}} 
                         - \frac{s_{13}}{s_{24}}- \frac{s_{13}}{s_{23}}\nonumber\\ 
C^{-+++-}_{F\,4b} &=& -\frac{s_{13}^{2}}{2s_{14}s_{34}}-\frac{(s_{13}+s_{15})^{2}}{2s_{14}s_{34}}+\frac{(s_{12}-s_{13})}{s_{34}}+\frac{s_{12}}{s_{23}}\nonumber
\end{eqnarray}
The nonvanishing contribution of the 2-point integrals can be cast into the form 
\begin{eqnarray}
{\cal A}^{-+++-}_B &=&
\Bigl\{ \Bigl[ C^{-+++-}_{B\,1}\;  p_1\cdot {\cal F}_2^+ \cdot p_3 \;I_2^n(s_{12}) \nonumber\\ 
&&   \;\; + C^{-+++-}_{B\,2}\;  p_1\cdot {\cal F}_2^+ \cdot p_3 \;I_2^n(s_{25}) \Bigr]
\frac{\mbox{Tr}({\cal F}_1^-{\cal F}_5^-)\mbox{Tr}({\cal F}_3^+{\cal F}_4^+)}{s_{15}^2 s_{34}^2}\nonumber\\   
&& + \Bigl[[ C^{-+++-}_{B\,3}\;  p_1\cdot {\cal F}_4^+ \cdot p_2  + ( 2 \leftrightarrow 3 ) ]\;I_2^n(s_{45})\nonumber\\  
&& \;\;+ [ C^{-+++-}_{B\,4a}\;  p_1\cdot {\cal F}_4^+ \cdot p_2 + C^{-+++-}_{B\,4b}\;  p_1\cdot {\cal F}_4^+ \cdot p_3 )]\;I_2^n(s_{14})\Bigr]
 \frac{\mbox{Tr}({\cal F}_1^-{\cal F}_5^-)\mbox{Tr}({\cal F}_2^+{\cal F}_3^+)}{s_{15}^2 s_{23}^2}\Bigl\}\nonumber\\
&& - ( 3 \leftrightarrow 4 )\nonumber
\end{eqnarray} 
Note that $I_2^n(s) \sim (-s)^{-\epsilon}/\epsilon$ leads to a logarithmic term and a pole part. The latter
has to vanish because of the finiteness of the amplitude. Replacing the $I_2^n$ integrals by the pole 
part only leads to a nontrivial relation between the coefficients. We have checked 
algebraically that the sum over the pole parts is indeed zero by using the flipping and remapping
for our field strength representation rules (\ref{perm_rule},\ref{flip_rule}). Of course this provides another
non-trivial check of the calculation.  
The coefficients of the logarithms are    

\newpage

\begin{eqnarray}
&& C^{-+++-}_{B\,1} =\nonumber \\ && 
\Bigl[
-\frac{2(s_{12}+s_{23})(-s_{15}+s_{13})}{s_{24}^{2}}
-\frac{s_{45}^{2}(-s_{35}+s_{24})}{s_{12}(s_{14}+s_{24})^{2}}
-\frac{s_{45}(s_{45}+2s_{24})}{s_{12}(s_{14}+s_{24})}
-\frac{2s_{14}(s_{14}-s_{24})}{s_{23}s_{45}}
\nonumber\\&&
-\frac{s_{14}^{2}(-s_{14}+4s_{12})}{s_{45}s_{23}^{2}}
+\frac{(s_{12}+s_{24})s_{24}^{2}}{s_{12}s_{23}^{2}}
+\frac{s_{14}(s_{23}+2s_{12}-2s_{24})}{s_{23}s_{45}}
-\frac{s_{23}(3s_{14}-s_{23}-2s_{45})}{s_{12}s_{24}}
\nonumber\\&&
-\frac{s_{24}(4s_{45}+s_{14}-2s_{24})}{s_{23}s_{12}}
-\frac{2s_{35}s_{24}}{s_{12}(s_{23}+s_{13})}
-\frac{3s_{24}s_{35}^{2}}{s_{12}s_{23}^{2}}
-\frac{(2s_{12}+s_{24})(-s_{24}+s_{13})^{2}}{s_{24}^{2}s_{35}}
\nonumber\\&&
-\frac{4s_{45}s_{14}}{s_{23}s_{12}}
-\frac{(s_{12}+s_{24})(-s_{24}+3s_{45}+8s_{14})}{s_{23}^{2}}
-\frac{s_{45}^{2}}{s_{12}s_{24}}
-\frac{s_{35}(s_{35}-s_{23})^{2}s_{45}^{2}}{s_{12}s_{23}^{2}(s_{23}+s_{13})^{2}}
\nonumber\\&&
-\frac{s_{24}s_{35}^{2}}{(s_{23}+s_{13})^{2}s_{12}}
-\frac{(2s_{12}+2s_{24}-s_{35})}{(s_{23}+s_{13})}
-\frac{(-2s_{45}+s_{34})}{s_{12}}
+\frac{(3s_{12}+3s_{15}+s_{23})}{s_{24}}
\nonumber\\&&
-\frac{s_{24}s_{35}s_{45}}{s_{12}(s_{14}+s_{24})^{2}}
+\frac{s_{45}s_{35}^{2}(3s_{24}+s_{35})}{s_{12}s_{23}^{2}(s_{23}+s_{13})}
+\frac{s_{45}^{2}s_{35}}{s_{24}s_{12}(s_{14}+s_{24})}
+\frac{s_{35}s_{45}s_{24}(-s_{23}+2s_{35})}{(s_{23}+s_{13})^{2}s_{12}s_{23}}
\nonumber\\&&
+\frac{2(s_{12}s_{34}+s_{12}s_{24}-s_{45}s_{24})^{3}}{s_{23}^{3}(s_{23}+s_{13})s_{12}s_{45}}
+\frac{(s_{12}+s_{24})s_{24}(2s_{35}-3s_{45})}{s_{12}s_{23}^{2}}+\frac{(2s_{12}+4s_{15}+4s_{24})}{s_{23}}
\nonumber\\&&
-\frac{2s_{35}(s_{24}+s_{45})(s_{35}-2s_{45})}{s_{12}s_{23}(s_{23}+s_{13})}
-\frac{(2s_{24}-2s_{45})}{(s_{14}+s_{24})}
-\frac{2s_{23}s_{35}(s_{12}+s_{23})}{s_{12}s_{24}^{2}}
\Bigr]
\nonumber\end{eqnarray}

\begin{eqnarray}
&& C^{-+++-}_{B\,2} =\nonumber \\ && 
\Bigl[
-\frac{s_{14}(s_{14}s_{25}+s_{12}s_{24})}{(s_{24}+s_{45})s_{12}s_{24}}+\frac{s_{14}}{s_{24}}
-\frac{s_{14}s_{24}s_{13}^{2}}{(s_{23}+s_{35})s_{23}^{2}s_{12}}-\frac{s_{24}s_{35}}{s_{23}^{2}}
+\frac{s_{24}s_{13}^{2}}{s_{12}s_{23}^{2}}-\frac{s_{25}(s_{13}+s_{35})^{2}}{s_{23}^{2}(s_{23}+s_{35})}\Bigr]
\nonumber\end{eqnarray}
\begin{eqnarray}
&& C^{-+++-}_{B\,3} =\nonumber \\ && 
\Bigl[
-\frac{s_{23}^{2}(s_{24}+s_{34})}{s_{14}(s_{23}+s_{13})^{2}}
-\frac{s_{23}(s_{23}+s_{34})}{(s_{23}+s_{13})^{2}}
-\frac{s_{12}s_{23}^{2}}{s_{14}s_{34}(s_{23}+s_{13})}
+\frac{s_{45}(s_{24}+s_{45})s_{23}^{2}}{(s_{12}+s_{23})^{2}s_{14}s_{24}}
\nonumber \\ &&
-\frac{s_{45}(s_{24}+2s_{23})}{(s_{12}+s_{23})^{2}}
-\frac{s_{12}^{2}s_{34}^{2}}{s_{14}s_{24}^{2}(s_{12}+s_{23})}
+\frac{(s_{24}+s_{45})s_{23}^{2}}{s_{14}s_{24}(s_{12}+s_{23})}
+\frac{s_{23}(2s_{34}s_{12}-s_{14}s_{23})}{s_{24}^{2}(s_{12}+s_{23})}
\nonumber \\ &&
-\frac{2s_{13}(s_{23}+s_{24})}{(s_{12}+s_{23})s_{23}}
-\frac{(s_{12}s_{24}+s_{23}^{2}+2s_{23}s_{24}+s_{24}^{2})}{s_{14}s_{34}}
+\frac{s_{34}(3s_{24}-3s_{35}+s_{23}-s_{34})}{s_{24}s_{14}}
\nonumber \\ &&
+\frac{(s_{24}+s_{14})}{(s_{12}+s_{23})}
+\frac{(2s_{23}+2s_{34})}{s_{24}}
+\frac{2s_{34}}{s_{23}}
-\frac{(s_{23}+3s_{34}-3s_{25})}{s_{14}}-\frac{(s_{23}+s_{24})}{s_{34}}
-\frac{s_{34}}{(s_{23}+s_{13})}\Bigr]
\nonumber\end{eqnarray}

\begin{eqnarray}
&& C^{-+++-}_{B\,4a} =\nonumber \\ && \Bigl[
\frac{s_{34}s_{23}^{2}}{s_{14}(s_{12}+s_{24})^{2}}
-\frac{s_{23}^{3}s_{14}}{s_{24}^{2}(s_{12}+s_{24})^{2}}
-\frac{s_{23}(s_{35}+s_{34})(s_{24}+2s_{23})}{s_{24}(s_{12}+s_{24})^{2}}
-\frac{s_{23}s_{34}}{s_{35}(s_{12}+s_{24})}
\nonumber \\ &&
+\frac{(s_{35}+s_{34})}{(s_{12}+s_{24})}
+\frac{s_{23}^{2}(3s_{13}+2s_{23})}{s_{24}^{2}(s_{12}+s_{24})}
-\frac{s_{35}s_{34}}{s_{14}(s_{23}+s_{35})}
-\frac{(s_{34}-s_{35})}{(s_{23}+s_{35})}
-\frac{3s_{34}(2s_{34}+s_{13})}{s_{24}s_{14}}
\nonumber \\ &&
-\frac{s_{34}^{2}(3s_{34}-3s_{35}+2s_{13})}{s_{24}^{2}s_{14}}
+\frac{2s_{34}^{3}s_{35}}{s_{14}s_{24}^{3}}
+\frac{(s_{13}+s_{34})s_{34}}{s_{35}s_{14}}
+\frac{s_{15}^{2}(s_{15}-s_{24})(2s_{14}+s_{24})}{s_{35}s_{24}^{3}}-\frac{3s_{15}^{2}}{s_{24}^{2}}
\nonumber \\ &&
-\frac{2s_{35}s_{23}^{3}}{s_{24}^{3}(s_{12}+s_{24})}
+\frac{2s_{15}^{2}(s_{15}-3s_{34})}{s_{24}^{3}}
-\frac{s_{24}(s_{23}+s_{35})}{s_{34}s_{14}}
-\frac{(s_{34}+s_{15}-s_{24})}{s_{34}}-\frac{(s_{24}+s_{34}+s_{35})}{s_{14}}\Bigr]
\nonumber\end{eqnarray}

\begin{eqnarray}
&& C^{-+++-}_{B\,4b} =\nonumber \\ && \Bigl[
-\frac{s_{24}s_{23}^{2}}{s_{35}(s_{12}+s_{24})^{2}}
+\frac{s_{23}(s_{23}+s_{24})}{(s_{12}+s_{24})^{2}}
-\frac{s_{24}s_{23}^{2}}{s_{14}s_{35}(s_{12}+s_{24})}
+\frac{(s_{34}-2s_{35})(s_{24}+s_{14})}{s_{34}(s_{23}+s_{35})}
\nonumber \\ &&
+\frac{(s_{23}-s_{24})(s_{23}+s_{24})}{s_{24}(s_{12}+s_{24})}
+\frac{(s_{24}s_{35}+s_{14}s_{23}+s_{23}s_{24})^{2}}{s_{34}^{2}s_{14}(s_{23}+s_{35})}
+\frac{s_{24}s_{35}}{s_{14}(s_{23}+s_{35})}
+\frac{s_{23}s_{24}}{s_{14}(s_{12}+s_{24})}
\nonumber \\ &&
-\frac{s_{15}^{2}(2s_{14}+s_{24})}{s_{24}^{2}s_{35}}+\frac{s_{24}(s_{35}+s_{24})}{s_{14}s_{34}}
-\frac{s_{34}(2s_{23}-s_{34}+5s_{35})}{s_{24}s_{14}}+\frac{(3s_{23}+9s_{34}+6s_{24})}{s_{24}}
\nonumber \\ &&
-\frac{2s_{34}^{2}s_{35}}{s_{24}^{2}s_{14}}+\frac{4s_{34}(s_{23}+s_{34})}{s_{24}^{2}}+\frac{(3s_{24}+2s_{14})}{s_{34}}
-\frac{(3s_{23}-3s_{34}+3s_{35}-2s_{24})}{s_{14}}\Bigr]
\nonumber\end{eqnarray}

Finally the constant term reads
\begin{eqnarray}
{\cal A}^{-+++-}_1 &=&
\frac{\mbox{Tr}({\cal F}_1^-{\cal F}_5^-)\mbox{Tr}({\cal F}_3^+{\cal F}_4^+)}{s_{15}^2 s_{34}^2}
[ C^{-+++-}_{1}\;  p_1\cdot {\cal F}_2^+ \cdot p_3  - ( 3 \leftrightarrow 4 ) ]\nonumber
\end{eqnarray} 

\begin{eqnarray}
&& C^{-+++-}_1 =\nonumber \\ && \Bigl[
-\frac{3s_{24}(2s_{24}-2s_{35}-s_{45})}{s_{23}^{2}}
+\frac{s_{12}s_{45}}{s_{23}^{2}}
-\frac{s_{12}(2s_{12}-3s_{35}+6s_{24})}{s_{23}^{2}}
-\frac{(s_{23}+s_{24})s_{35}s_{23}}{s_{24}^{2}(s_{12}+s_{24})}
\nonumber \\ && 
-\frac{s_{45}^{2}s_{24}}{s_{12}s_{35}(s_{34}+s_{45})}
+\frac{s_{24}(s_{23}+s_{35})^{2}}{s_{45}s_{12}(s_{12}+s_{23})}
+\frac{s_{24}^{3}s_{45}}{s_{12}s_{23}^{2}(s_{12}+s_{23})}
+\frac{s_{24}^{2}(s_{24}-3s_{35})}{s_{12}s_{23}(s_{12}+s_{23})}
\nonumber \\ && 
-\frac{s_{24}(3s_{45}s_{24}+3s_{45}s_{35}+2s_{35}^{2})}{s_{12}s_{45}s_{23}}
-\frac{2s_{35}^{2}s_{24}}{s_{23}(s_{23}+s_{13})s_{45}}
+\frac{(s_{23}+s_{24}+s_{35})}{(s_{12}+s_{23})}
+\frac{s_{35}^{2}s_{24}}{s_{12}s_{45}^{2}}
\nonumber \\ &&
+\frac{(s_{35}+2s_{13}-2s_{23})}{s_{24}}
-\frac{(2s_{24}-2s_{25})}{s_{23}}
+\frac{(s_{23}+s_{24})(s_{23}-s_{24}-2s_{45})}{s_{12}(s_{12}+s_{24})}
-\frac{s_{35}(s_{23}+s_{24})}{s_{23}(s_{12}+s_{24})}
\nonumber \\ &&
-\frac{s_{45}(s_{24}-s_{45})}{s_{35}(s_{34}+s_{45})}
+\frac{s_{24}(s_{24}+2s_{35})}{s_{23}(s_{12}+s_{23})}
+\frac{s_{24}s_{35}}{s_{45}(s_{23}+s_{13})}
+\frac{s_{35}^{2}s_{24}}{s_{45}^{2}(s_{23}+s_{13})}
-\frac{s_{24}(s_{23}-s_{35})}{s_{12}s_{45}}
\nonumber \\ &&
+\frac{s_{35}(s_{23}-s_{35})^{2}}{s_{23}^{2}(s_{23}+s_{13})}
-\frac{s_{24}^{2}(2s_{24}-3s_{35}-3s_{45})}{s_{12}s_{23}^{2}}
-\frac{s_{23}(-2s_{45}-3s_{35}+2s_{23})}{s_{24}s_{12}}
\nonumber \\ &&
+\frac{s_{23}^{2}s_{35}}{s_{12}s_{24}^{2}}
+\frac{s_{45}(3s_{24}-s_{45})}{s_{12}s_{35}}
+\frac{s_{13}(s_{13}-2s_{24})}{s_{24}s_{35}}
-\frac{(-s_{24}+s_{45}+3s_{23})}{s_{12}}
-\frac{s_{14}(-s_{23}+s_{14})^{2}}{s_{45}s_{23}^{2}}\Bigr]
\nonumber\end{eqnarray}

\section{Summary}
We have presented a calculation of the one-loop amplitude $\gamma\gamma ggg\to 0$.
By applying the spinor helicity method a decomposition in helicity amplitudes was obtained
which in the end were reexpressed in terms of field strength tensors. 
We have derived useful relations for invariants built out of field strength tensors
to switch between different field strength representations.
This approach  avoids complex-valued spinor products in the numerical evaluation of cross sections.  
Two independent  algebraic programs have been designed with different implementations
of the spinor helicity framework. We find agreement of both results.
Further we have checked the  Bose symmetries of the given helicity amplitudes.
The Ward identities are manifestly fulfilled in the given representation.
 
The decomposition of the algebraic expressions in terms of  an appropriate function basis 
leads to a very efficient organization of the calculation which is fully automated by using algebraic programs.  
The final result is compact which suggests that our method is well designed to deal with more
complex amplitudes.  

A phenomenological application of our calculation will be provided elsewhere.

\section*{Acknowledgment}

This work was supported by the Bundesministerium f\"ur Bildung und
Forschung (BMBF, Bonn, Germany) under the contract number 05HT4WWA2.

\end{document}